\documentclass{iopart} 

\usepackage{graphicx}
\usepackage{fixltx2e}

\usepackage[breaklinks]{hyperref}
\hypersetup{
    bookmarks=false,
    pdfstartview={FitH},
    colorlinks=true,
    linkcolor=blue,
    citecolor=blue,
    urlcolor=blue
}

\begin{document}

\title{Low-energy V {\em t}\textsubscript{2{\em g}} orbital
excitations in NdVO\textsubscript{3}}

\author{J~Laverock$^1$, B~Chen$^1$, A~R~H~Preston$^1$,
D~Newby,~Jr$^1$, L~F~J~Piper$^{1,2}$,
L~D~Tung$^3$\footnote{Present address:
Department of Physics \& Astronomy, University College London, Gower Street,
London WC1E 6BT, UK},
G~Balakrishnan$^3$, P-A~Glans$^4$, J-H~Guo$^4$ and
K~E~Smith$^{1,5}$}

\address{$^1$ Department of Physics, Boston University, 590 Commonwealth Avenue,
Boston, MA 02215, USA}
\address{$^2$ Department of Physics, Applied Physics and Astronomy, Binghamton
University, Binghamton, NY 13902, USA}
\address{$^3$ Department of Physics, University of Warwick, Coventry, CV4 7AL,
United Kingdom}
\address{$^4$ Advanced Light Source, Lawrence Berkeley National Laboratory,
Berkeley, CA 94720, USA}
\address{$^5$ School of Chemical Sciences and The MacDiarmid Institute for
Advanced Materials and Nanotechnology, The University of Auckland, Private
Bag 92019, Auckland 1142, New Zealand}

\ead{laverock@bu.edu}

\pacs{78.70.Ck,75.25.Dk,73.20.Mf}

\begin{abstract}
The electronic structure of NdVO$_3$ and YVO$_3$ has been investigated
as a function of sample temperature using resonant inelastic soft x-ray
scattering at the V $L_3$-edge. Most of the observed spectral features are
in good agreement with an atomic crystal-field multiplet model. However, a
low energy feature is observed at $\sim 0.4$~eV that cannot be explained by
crystal-field arguments. The resonant behaviour of this feature establishes
it as due to excitations of the V $t_{2g}$ states. Moreover, this feature
exhibits a strong sample temperature dependence, reaching maximum intensity
in the orbitally-ordered phase of NdVO$_3$, before becoming suppressed at
low temperatures. This behaviour indicates that the origin of this feature
is a collective orbital excitation, i.e.~the bi-orbiton.
\end{abstract}

\maketitle

\section{Introduction}
The competition and interplay between spin, orbital and charge degrees
of freedom in transition metal oxides yield extremely rich, and often
unexpected, phase diagrams\cite{tokura2000etc}. In contrast to $e_g$
active systems, such as the manganites, partially filled $t_{2g}$ electron
systems experience weaker coupling to the lattice, leading to weaker
Jahn-Teller distortions. Alongside the strong coupling between spin and
orbital degrees of freedom in $t_{2g}$ systems, this leads to a wealth
of near-degenerate spin- orbital- and charge-ordered structures that
are relevant at similar temperatures. In particular, in the rare-earth
orthovanadates, {\em R}VO$_3$ ({\em R}~=~Y, La~--~Lu), both orbital order
(OO) and antiferromagnetic spin order (SO) develop in at least three distinct
ordered phases, concomitant with structural transitions and Jahn-Teller
distortions\cite{blake2001etc,miyasaka2003}.

Below the OO transition temperature ($T_{\rm OO} \sim 200$~K), the V $t_{2g}$
electrons in {\em R}VO$_3$
become orbitally ordered in a G-type (G-OO) arrangement\cite{miyasaka2003},
in which $d_{xy}$ orbitals are always occupied and the $d_{yz}$ and $d_{zx}$
orbitals are alternately occupied. Below the N\'{e}el temperature,
$T_N \sim 120$~K, C-type SO
(C-SO) sets in, retaining the G-OO, and in which the spins are aligned
antiferromagnetically in the $ab$ plane, and ferromagnetically along $c$. For
the smaller rare-earth ions (Y, Dy -- Lu), an additional transition occurs
below $\sim 80$~K, in which the OO switches to C-type (C-OO) and the SO to
G-type (G-SO)\cite{miyasaka2003}. More recently, there has been some evidence
that C-OO coexists with G-OO below $T_N$ for some of the larger rare-earths,
including NdVO$_3$, leading to phase-separated regions that grow from small
droplets\cite{sage2006,sage2007}. Optical measurements have observed a
substantial temperature dependence in optically-allowed $dd^*$ excitations,
at energies of 1.8, 2.4 and 3.3~eV\cite{tsvetkov2004,reul2012}. Although
on-site $dd^*$ transitions are optically forbidden, dimer-type excitations,
$d^2\,d^2 \rightarrow d^3\,d^1$, are allowed, and the optical excitations have
been interpreted in terms of the $d^3$ multiplet structure\cite{reul2012}.
Subsequent time domain spectroscopy measurements have suggested the low-energy
excitation is a Hubbard exciton, i.e.~a bound exciton across the Mott-Hubbard
gap\cite{novelli2012}. At lower energies, optical conductivity measurements
have found a peak at $\sim 0.4$~eV, i.e.~within the Mott-Hubbard gap, which
the authors attribute to a collective orbital excitation\cite{saitoh2001}
in the form of a
bi-orbiton\cite{benckiser2008}. Although previous orbiton assignments of
optical data in {\em R}VO$_3$ have been controversial (see, for example,
Refs.~\cite{miyasaka2005etc,sugai2006}), such excitations have
previously been observed at similar energies in the $d^1$ titanates with
both optical spectroscopy\cite{ulrich2006} and resonant inelastic x-ray
scattering (RIXS)\cite{ulrich2009}. In the cuprates, the momentum dispersion
of the orbiton has recently been revealed by high-resolution RIXS, providing
unambiguous evidence of their origin\cite{schlappa2012}.

RIXS is fast developing as a powerful and valuable probe of correlated electron
systems (e.g.~Refs.~\cite{schlappa2012,braicovich2009etc}).
Transition metal $L$-edge RIXS is a bulk sensitive, direct two-step
process\cite{ament2011b}, in which the incident photon is resonantly
tuned to a feature of the absorption spectrum, i.e.~a core level transition,
leading to the unambiguous association of RIXS features with a particular
atomic site and even a particular
orbital. The specific direct RIXS process at the V $L$-edge in
{\em R}VO$_3$ can be summarised as, $2p^6\,3d^2 \rightarrow 2p^5\,3d^3
\rightarrow 2p^6\,3d^{2*}$, where the $^*$ in the final state indicates it may
be an excited state or configuration of the ion. For example, $dd^*$ transitions
include excitations of the $3d$ electrons both within the V $t_{2g}$ manifold,
and between $t_{2g}$ and $e_g$ states. Moreover, unlike optical measurements,
soft
x-rays carry appreciable momentum, and can probe excitations away from the
zone centre. Here, we present a detailed study of the RIXS excitations of the
V $t_{2g}$ electrons of NdVO$_3$, demonstrating that a low energy feature
at $\sim 0.4$~eV may be associated with the delocalised orbital excitation
of the V $t_{2g}$ electrons.

\section{Methods}
High-quality single crystals of NdVO$_3$ and YVO$_3$ were grown by the
floating zone technique as described in Ref.~\cite{tung2007etc}.
Soft x-ray spectroscopy measurements were carried out at beamline X1B of
the National Synchrotron Light Source, Brookhaven and the AXIS endstation of
beamline 7.0.1 at the Advanced Light Source, Berkeley. Samples were cleaved
{\em ex-situ}, and loaded into ultra-high vacuum within 10 minutes to ensure
fresh, clean surfaces. X-ray absorption spectroscopy (XAS) measurements were
performed in total electron yield (TEY) mode with an incident photon energy
resolution ($\Delta E_{\rm in}$) of 0.2~eV at FWHM, and the photon energy
was calibrated using TiO$_2$ reference spectra of the Ti $L$-edge and O
$K$-edge. RIXS spectra were recorded with a Nordgren-type
spectrometer\cite{nordgren1986}, and
the instrument was calibrated using a Zn reference spectrum. The incident
photon was polarised in the scattering plane ($\pi$ scattering geometry),
and the scattering angle was 90$^{\circ}$, with the surface normal at
45$^{\circ}$ to the incident photons.  All spectra used to construct the RIXS
intensity map of NdVO$_3$ were recorded during the same sample conditions
with spectrometer resolution ($\Delta E_{\rm spec}$) and $\Delta E_{\rm in}$
of 0.36~eV at FWHM\cite{laverock2011b}.
Additional, separate, measurements on NdVO$_3$ and YVO$_3$
were performed at several temperatures between 80~K and room temperature at
higher resolution, with $\Delta E_{\rm spec} = 0.26$~eV and $\Delta E_{\rm
in} = 0.21$~eV (with combined resolution $\approx 0.33$~eV). In these
measurements, the incident photon energy was kept
fixed to ensure quantitative comparison between spectra. Since the natural pixel
width of these measurements is just one-half of the FWHM of the resolution
function, we have employed sub-pixel sampling to these higher resolution
measurements (see Appendix for details). For presentation purposes, and since
error bars are difficult to quantify in this process, we show error bars at the
natural pixel width of the measurement, with the sub-pixel sampled data shown as
a `guide for the eye'.

\begin{figure}[t!]
\begin{center}
\includegraphics[width=0.6\linewidth,clip]{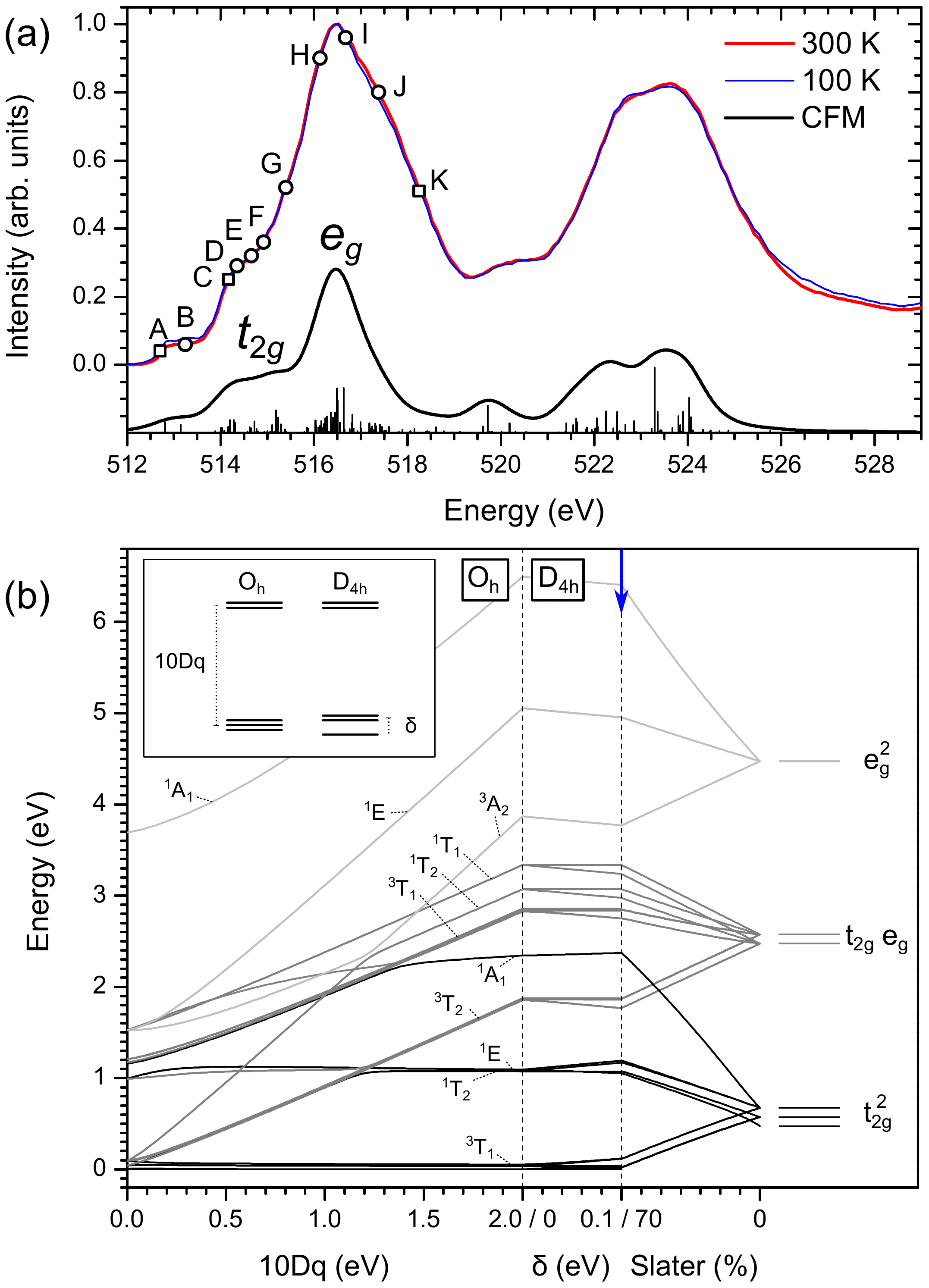}
\end{center}
\vspace*{-0.2in}
\caption{\label{f:xas} (a) V $L_{3,2}$-edge XAS spectra of NdVO$_3$ at room
temperature and 100~K, compared with the atomic CFM model ($10Dq = 2.0$~eV,
$\delta = 0.1$~eV, and Slater reduction of 70\%). (b)
Atomic multiplet energy level diagram for the V$^{3+}$ $d^2$ ion in $O_h$
symmetry for $0 \leq 10Dq \leq 2.0$~eV, followed by their splitting with
$0 \leq \delta \leq 0.1$~eV in $D_{4h}$ symmetry. The high-field
parentage is shown on the right, corresponding to the reduction of the Slater
integrals from $70\% \rightarrow 0$. The inset schematically illustrates
the CF energies, and definition of $\delta$.}
\end{figure}

\section{Results}
V $L_{3,2}$-edge XAS spectra are shown in Fig.~\ref{f:xas}(a), and are in
very good agreement with similar measurements of
YVO$_3$\cite{pen1999,benckiser2013}, as
well as with atomic crystal field multiplet (CFM) calculations of the $d^2$
V$^{3+}$ ion, shown in tetragonal $D_{4h}$ symmetry at the bottom of the
figure. Since the $e_g$ states are well separated from the $t_{2g}$ states
in {\em R}VO$_3$, we choose combinations of the crystal field (CF) parameters
$D_s$ and $D_t$ that preserve the degeneracy of the $e_g$ levels, leaving
just a single parameter that characterises the lowering in symmetry between
$O_h$ and $D_{4h}$. This parameter, $\delta$ ($D_s = -\delta/7$ and $D_t =
4\delta/35$), represents the splitting between the $d_{xy}$ and $d_{xz,yz}$
states, and is illustrated in the inset to Fig.~\ref{f:xas}(b).  The CFM
spectrum in Fig.~\ref{f:xas}(a) is calculated with $10Dq = 2.0$~eV and
$\delta = 0.1$~eV, approximately accounting for the lowering in energy of
the $d_{xy}$ level by 0.1~--~0.2~eV that is anticipated by first-principles
studies\cite{otsuka2006,deraychaudhury2007} and which has recently been observed
directly in
high-resolution RIXS measurements of YVO$_3$\cite{benckiser2013}. From the
CFM calculations, we
associate the shoulder around 515~eV predominantly with excitations within
the $t_{2g}$ states, and the peak at 516.5~eV with excitations into the $e_g$
states. The experimental XAS spectra are found to be relatively insensitive to
temperature; for example, the room temperature spectra are slightly broader,
with a larger contribution $\sim 517.5$~eV and a weaker intensity of the
pre-edge features at $\sim 513$~eV. We associate these very slight differences
(which are reversible and reproducible) with the small change in the local
CF between the high temperature orthorhombic phase and the low temperature
monoclinic phase\cite{deraychaudhury2007}.

\begin{figure}[t!]
\begin{center}
\includegraphics[width=0.7\linewidth,clip]{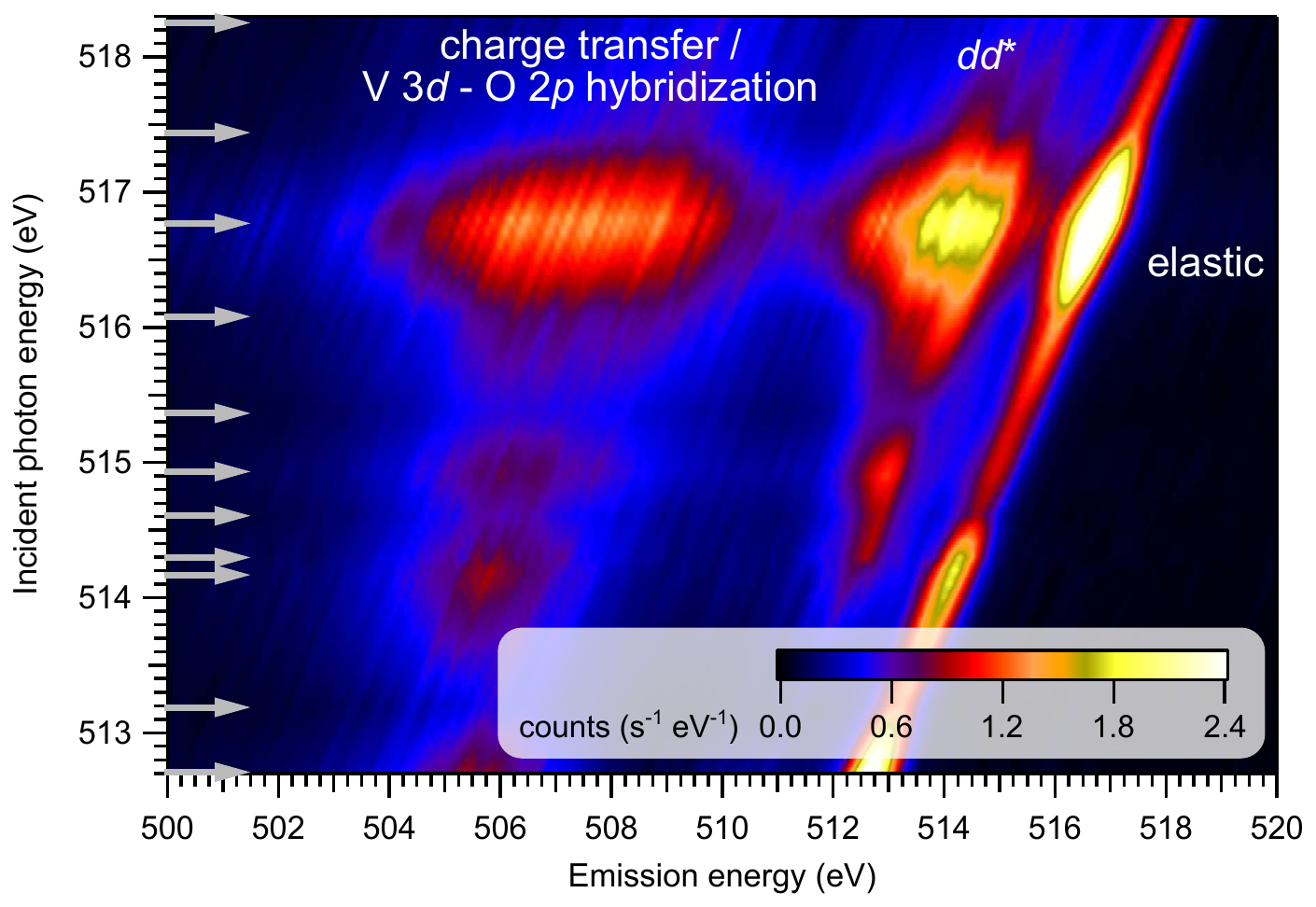}
\end{center}
\vspace*{-0.2in}
\caption{\label{f:rixsmap} RIXS intensity map of NdVO$_3$ at
room temperature.  The discrete incident photon energies used to construct
the intensity maps are shown by the arrows at the left of the figure. The
colour intensity scale represents the experimental transition rate.}
\end{figure}

In order to separate localised CF ($dd^*$) transitions from delocalised
excitations in the RIXS data, calculations of the multiplet configurations
have been performed in both octahedral ($O_h$) and $D_{4h}$ symmetry,
shown in Fig.~\ref{f:xas}(b).  In wide band-gap insulators (such as {\em
R}VO$_3$), local CF ($dd^*$) transitions in RIXS are described well by the
CFM model\cite{degroot2005etc,laverock2013}. On the left of
Fig.~\ref{f:xas}(b), the evolution in multiplet energies are shown as $10Dq$
is increased from zero (the free ion picture) to 2.0~eV in $O_h$, including
spin-orbit coupling (which weakly splits some of the lines). Subsequently,
the $\delta$ parameter is increased in $D_{4h}$ symmetry to 0.1~eV,
approximating the CF of {\em R}VO$_3$ [and indicated by the arrow in
Fig.~\ref{f:xas}(b)]. Finally, the Slater integrals are reduced to zero,
reflecting the single-particle parentage of the states (i.e.~$t_{2g}^2$,
$t_{2g}^1\,e_g^1$ and $e_g^2$). In this picture, the $t_{2g}^2$ occupation
is further split into three levels: $d_{xy}^2$, $d_{xy}^1\,d_{xz,yz}^1$
and $d_{xz,yz}^2$.

The V $L_3$-edge RIXS measurements are summarised in Fig.~\ref{f:rixsmap},
shown as an intensity map constructed from eleven discrete incident photon
energies between 512.5 and 518.5~eV [indicated in Fig.~\ref{f:xas}(a)].
Below resonance, the spectra consist of quasi-elastic scattering (including
very low energy phonon contributions), with a weak contribution from
charge-transfer processes centered $\approx 7$~eV lower in energy, in
good agreement with estimations based on our photoemission measurements
(not shown).  As the incident photon energy is increased, the inelastic RIXS
processes (i.e.~$dd^*$ excitations) become more intense, developing into local
maxima at $\sim 514.5$~eV and $\sim 516.5$~eV, corresponding to excitations
into unoccupied $t_{2g}$ and $e_g$ states, respectively [as illustrated in
Fig.~\ref{f:xas}(a)].  Therefore, we can associate $dd^*$ loss features at the
first of these resonances (near 514.5~eV) predominantly with excitations within
the $t_{2g}$ manifold, whereas features at the second resonance (near 516.5~eV)
are associated to a greater extent with excitations into the $e_g$ states.

\begin{figure}[t!]
\begin{center}
\includegraphics[width=0.6\linewidth,clip]{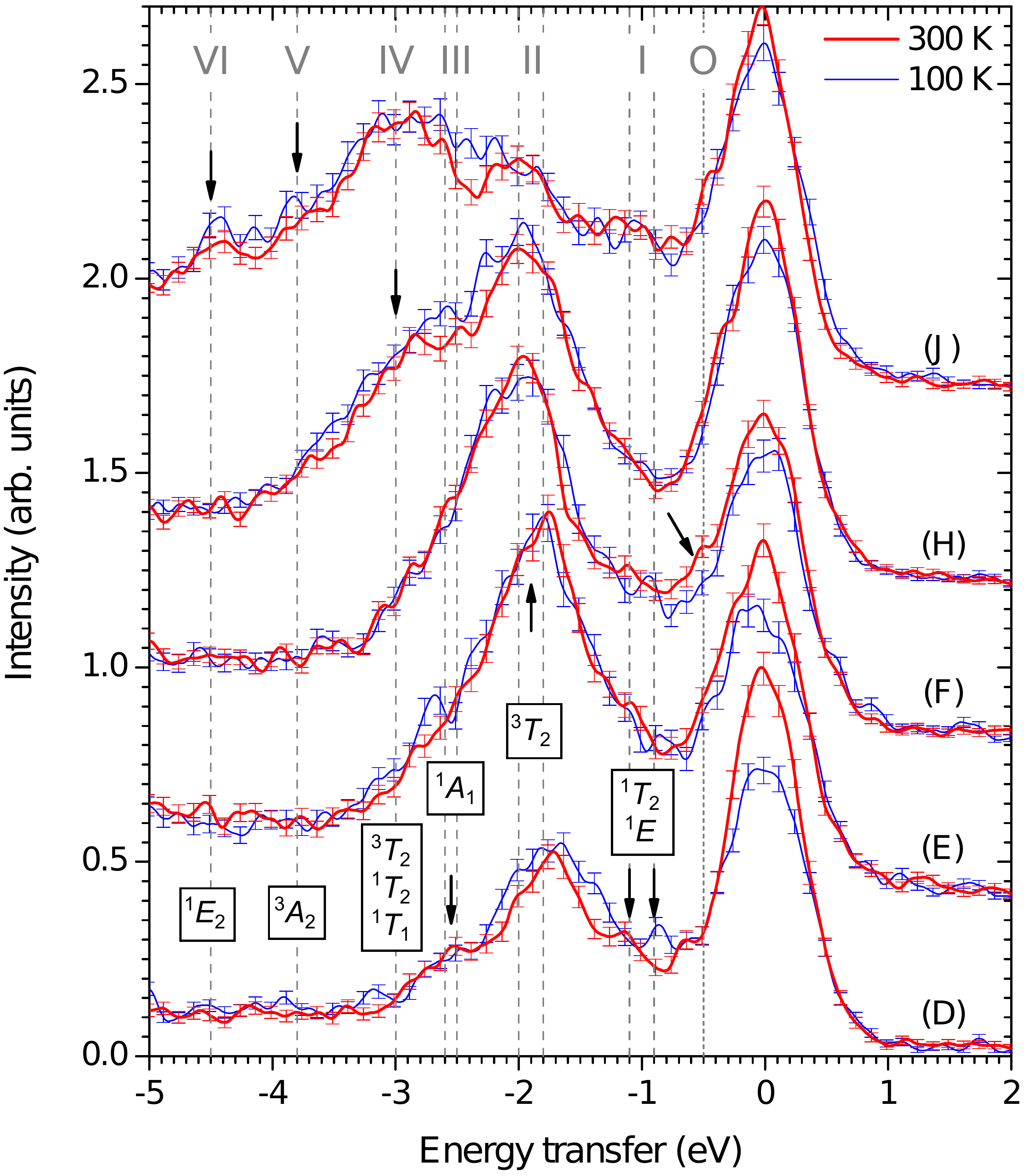}
\end{center}
\vspace*{-0.2in}
\caption{\label{f:rixs_max} Representative RIXS spectra
at room temperature and 100~K.}
\end{figure}

In Fig.~\ref{f:rixs_max}, five representative spectra are shown on
an energy transfer scale at room temperature and at 100~K, and are broadly
in good agreement with the recent high-resolution RIXS data of
Ref.~\cite{benckiser2013} on YVO$_3$.
The excitation energies used
for the two different temperatures have been carefully checked to agree to
within $\pm 0.1$~eV. Spectra (D-F), separated by just 0.6~eV, correspond to
energies at the $t_{2g}$ absorption resonance, whereas spectra (H) and (J)
are recorded near the $e_g$ resonance.  Note that there is only a very weak
dependence of the intensity of the $dd^*$ transitions with temperature between
the disordered phase (300~K) and spin- and orbitally-ordered (G-OO, C-SO)
phase at 100~K.  Through comparison with the CFM model of Fig.~\ref{f:xas}(b),
features I and III can be associated with singlet $t_{2g}^2$ excitations,
and II is a triplet transition into a $t_{2g}^1\,e_g^1$ excited state. At
higher energies, IV represents a broad group of $t_{2g}^1\,e_g^1$
transitions, and V and VI are $e_g^2$ final states. The symmetries of these
excitations (in $O_h$ and neglecting spin-orbit coupling) are labelled in
Fig.~\ref{f:rixs_max}. We emphasise that the $^1T_2$/$^1E$ transition in
Fig.~\ref{f:xas}(b) is relatively insensitive to the CF, and is primarily
determined by $2J_H = 2(3B+C)$\cite{tanabe1954}, where $J_H$ is the on-site
Hund's exchange parameter, and $B$ and $C$ are the Racah parameters. This
transition can therefore be used as an independent measurement of $J_H$
and $B$, irrespective of the agreement of the CFM model elsewhere,
yielding $J_H = 0.55$~eV and $B = 0.083$~eV, which are similar to, but
smaller than, typical values used in theoretical models of the orbital
order\cite{khaliullin2001etc,deraychaudhury2007}.

\begin{figure}[t!]
\begin{center}
\includegraphics[width=0.6\linewidth,clip]{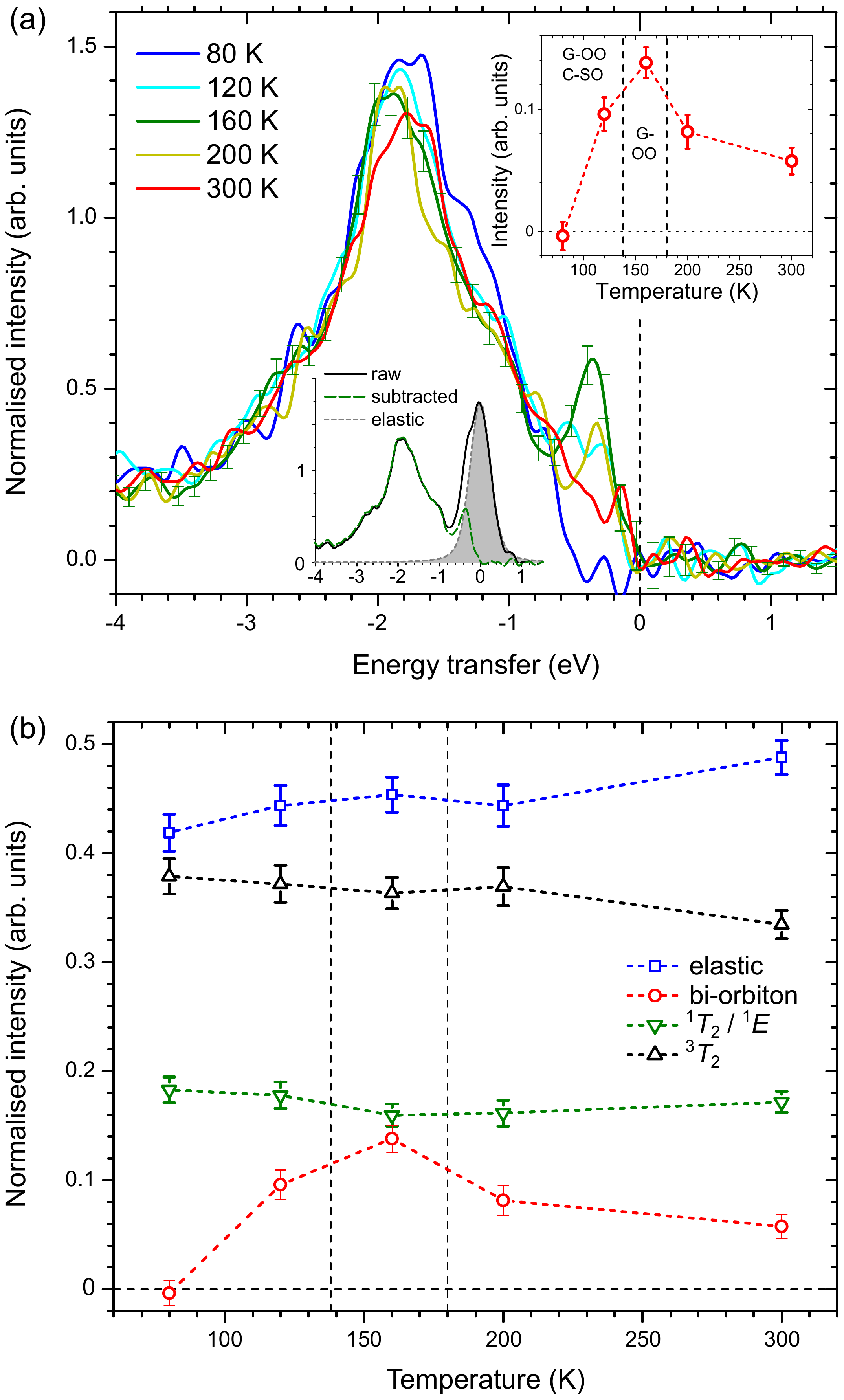}
\end{center}
\vspace*{-0.2in}
\caption{\label{f:ndtemper} Temperature dependence of
RIXS excitations in NdVO$_3$, recorded with an incident photon energy of
514.5~eV. (a) Spectra after removal of the elastic peak, showing the peak at
$-0.4$~eV that cannot be explained by the CFM model. Representative error bars
are shown for one of the spectra. An example of the elastic
peak subtraction is shown in the lower inset. The upper inset shows the
integrated intensity of the spectra between $-0.55$ and $-0.25$~eV.
(b) Temperature dependence of the RIXS excitations in (a). The intensity is
calculated in a 0.3~eV window centered about 0~eV (elastic peak), $-0.4$~eV
(bi-orbiton, after elastic peak subtraction),
$-1.0$~eV ($^1T_2$/$^1E$) and $-1.9$~eV ($^3T_2$).
The horizontal dotted line corresponds to zero intensity, and the vertical
dashed lines indicate $T_{\rm OO}$ and $T_N$\cite{miyasaka2003}.}
\end{figure}

Having established the origin of the CF excitations in the RIXS spectra, we
now turn to the feature that {\em cannot} be explained by the CFM model. At
$-0.4$~eV, a weak feature `O' is present for several different incident
energies at room temperature. This feature is found to resonate (have largest
intensity) at the $t_{2g}$ absorption resonance, indicating its origin
is excitations of the V $t_{2g}$ states, and is most apparent in spectrum
(F). At 100~K, it becomes indistinguishable from elastic scattering. This
feature is not expected from the CFM model, and we now investigate its
origin through higher resolution ($\approx 0.33$~eV combined resolution),
temperature-dependent measurements.
In Fig.~\ref{f:ndtemper}(a), five spectra between room temperature and 80~K are
shown at a single incident photon energy of 514.5~eV. This incident photon
energy [close to spectrum (E) in Fig.~\ref{f:rixs_max}]
is chosen to maximise the
contribution from `O' (resonant with $t_{2g}$ absorption). The quasi-elastic
scattering is strong, even at this energy, and partially obscures `O' in
its tails, although some asymmetry of this feature is clear, particularly at
160~K. In order to clarify the temperature behaviour of `O', the quasi-elastic
peak has been subtracted from the spectra in Fig.~\ref{f:ndtemper}(a)
[an example
of this subtraction is shown in the lower inset to
Fig.~\ref{f:ndtemper}(a)]. Owing
to contribution from other low-energy loss processes, such as phonons and
low-energy CF excitations\cite{benckiser2013}, the
elastic signal is fitted to a Voigt lineshape only above 0~eV, corresponding
only to instrument-broadened elastic scattering.
In Fig.~\ref{f:ndtemper}(a), the linewdith of the quasi-elastic peak has been
fitted to each spectrum independently, accounting for the temperature-dependent
evolution in the phonon contribution and thermal population of the CF ground
state. However, we emphasise that the following conclusions are insensitive to
such details, and we obtain similar results if a single (mean) linewidth is
used.
The subtracted spectra contain
a large temperature-dependent component centered at $-0.4$, which peaks at
160~K, before diminishing to zero at 80~K. The inset to Fig.~\ref{f:ndtemper}(a)
displays the integrated intensity between $-0.55$ and $-0.25$~eV of each
spectrum,
illustrating the large relative contribution of this feature in the OO phase,
before its suppression in the SO/OO phase. Even at room temperature, this
component has a reasonably large contribution. Although low-energy local CF
excitations are expected\cite{deraychaudhury2007} below $-0.2$~eV, they are
observed to be relatively weak, compared with elastic scattering, in YVO$_3$,
and do not display appreciable temperature dependence\cite{benckiser2013}.
Moreover, most of the spectral weight that contributes to the temperature
dependence observed in Fig.~\ref{f:ndtemper}(a) is concentrated well above
$-0.2$~eV. In Fig.~\ref{f:ndtemper}(b), this temperature dependence is compared
with the temperature dependence of elastic scattering and local CF excitations.
As anticipated by Fig.~\ref{f:rixs_max}, the elastic scattering and local CF
transitions above 80~K
are relatively insensitive to the orbital and spin order, showing
only weak (and approximately linear) variations in intensity with temperature,
in contrast to the pronounced temperature dependence of the $-0.4$~eV feature.
As we discuss in more detail below, the spectrum recorded at 80~K exhibits quite
different CF transition probabilities to the other spectra, with more weight
appearing at $-1.7$ and $-1.2$~eV.

\begin{figure}[t!]
\begin{center}
\includegraphics[width=0.45\linewidth,clip]{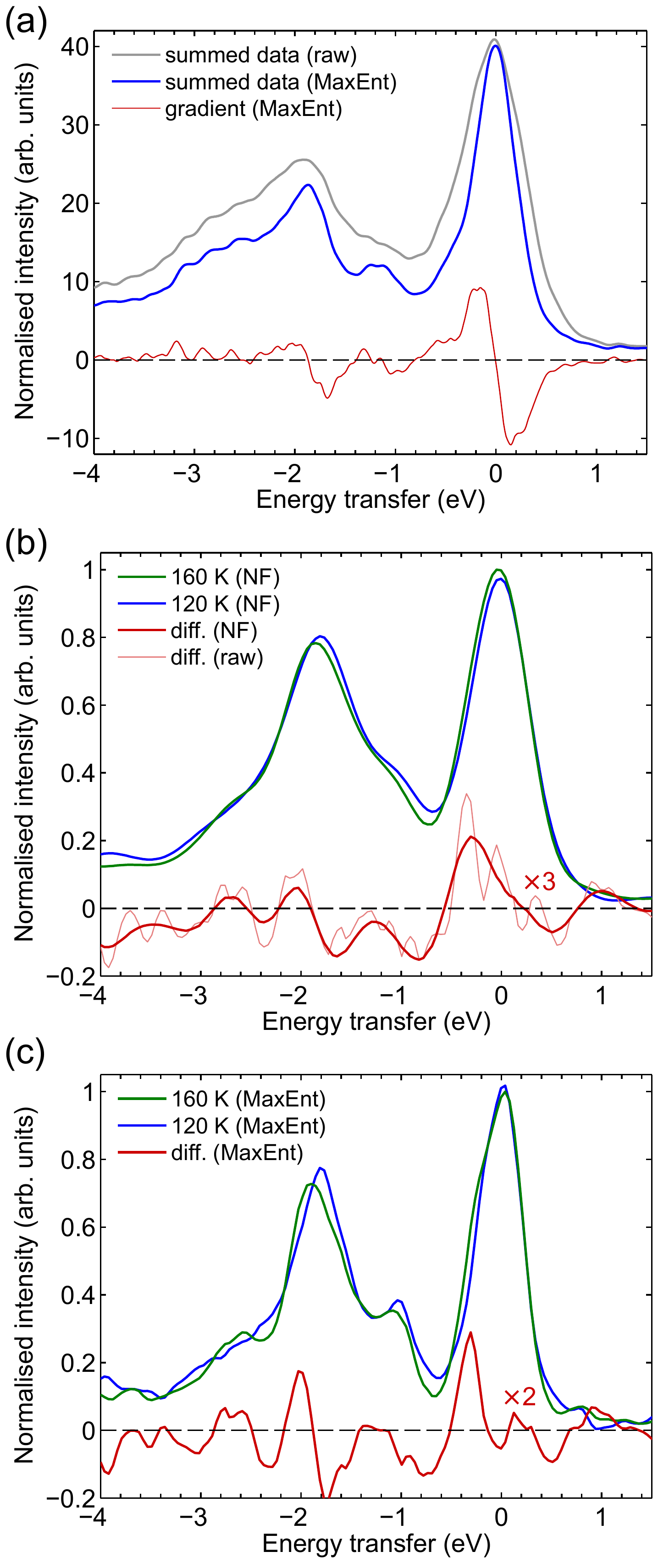}
\end{center}
\vspace*{-0.2in}
\caption{\label{f:anal} Parameter-free analysis of V $L_3$-edge RIXS spectra of
NdVO$_3$ showing the presence of a weak feature at $\approx -0.4$~eV. (a)~Summed
spectra at room temperature, obtained by integrating the RIXS map of
Fig.~\ref{f:rixsmap} between 514.7 and 517.3~eV, i.e.~over the $t_{2g}$ and
$e_g$ resonances. The summed spectrum has also been deconvoluted using the
MaxEnt procedure, and the gradient of this spectrum is also shown.
(b)~Difference between the 160~K and 120~K spectra of Fig.~\ref{f:ndtemper}
after applying a noise filter (NF) to the data. For comparison, the difference
between the raw spectra is also shown. (c)~The same 160~K and 120~K spectra
after being treated by the MaxEnt deconvolution procedure. Their difference is
also shown for comparison.}
\end{figure}

Before we discuss the origin of the low-energy excitation in more detail, we
briefly reinforce its presence through non-parameterised analysis. In
Fig.~\ref{f:anal}(a), the RIXS map of Fig.~\ref{f:rixsmap} has been integrated
over the $t_{2g}$ and $e_g$ resonances, yielding a (weighted) summed spectrum on
an energy transfer scale. Here, fluorescent features (which are dispersive in
energy transfer) are spread out and contribute a broad `background', whereas
RIXS features reinforce. A slight asymmetry of the elastic peak in
Fig.~\ref{f:anal}(a) develops into a prominent shoulder at $-0.4$~eV
after treatment with a
maximum-entropy deconvolution technique (MaxEnt)\cite{laverock2011b}. At the
bottom of Fig.~\ref{f:anal}(a), the gradient of this MaxEnt spectrum is shown,
in which the shoulder centered at $-0.4$~eV corresponds to feature `O'.

We now focus on the temperature-dependent data of Fig.~\ref{f:ndtemper}, and
employ difference spectra to examine the low-energy feature. Since the
quasi-elastic peak intensity evolves with temperature, we choose two
temperatures that are close together (120 and 160~K) to reduce problems
associated with the variation of the CF population with temperature. Owing to
the increased noise level of difference spectra, the raw spectra of
Fig.~\ref{f:ndtemper} are first passed through a noise filter\cite{laverock2013}
before subtraction, and are shown in Fig.~\ref{f:anal}(b). The difference
spectrum is
shown at the bottom of Fig.~\ref{f:anal}(b), illustrating the excess intensity
in the 160~K spectrum at $-0.4$~eV. For comparison, the difference in the raw
spectra is also shown, exhibiting the same structure. In Fig.~\ref{f:anal}(c),
the same two spectra have been treated with the MaxEnt deconvolution procedure,
and the extra spectral weight at $-0.4$~eV can be directly seen.
Finally, we note that the data presented in Fig.~\ref{f:anal}(a) and
Figs.~\ref{f:anal}(b-c) were recorded several months apart on different
cleaves of the sample boule, illustrating the reproducibility of the data.

\section{Discussion}
The above results indicate that the $-0.4$~eV
feature is associated with the orbital
order, but is strongly suppressed at low temperature. CF (multiplet)
excitations can be ruled out: low-energy multiplet excitations occur
on the order of the parameter $\delta$ [illustrated in the inset
to Fig.~\ref{f:xas}(b)], and are not expected as large as 0.4~eV.
In {\em R}TiO$_3$, whose CF splitting is roughly twice as large as the
vanadates\cite{deraychaudhury2007,pavarini2005}, local CF excitations
were predicted below 300~meV\cite{ulrich2009}. Moreover, it is hard to
reconcile the temperature evolution shown in Fig.~\ref{f:ndtemper}(b)
with local CF excitations. Secondly, the lowest energy dimer excitations
are expected at $3J_H$\cite{oles2005}, and are observed in optical spectra
at 1.8~eV\cite{tsvetkov2004}, and can therefore also be ruled out. Finally,
the energy is well above that of phonon and magnon modes, which occur below
100~meV in {\em R}VO$_3$\cite{miyasaka2005etc}.

Orbital excitations (in the form of bi-orbitons) have previously been observed
with RIXS at the Ti $L_3$-edge of LaTiO$_3$ and YTiO$_3$ at an energy of
0.25~eV\cite{ulrich2009}, following an earlier observation of a similar
feature in optical measurements\cite{ulrich2006}. These excitations were
described well by a superexchange model, in which the bi-orbiton process
(which displays only very weak dispersion) was found to be of stronger
intensity in the RIXS spectrum than single orbiton modes.  In YVO$_3$
and HoVO$_3$, optical measurements exhibit a strong peak at 0.4~eV, which
the authors attribute to the collective bi-orbiton\cite{benckiser2008}.
This feature, present at room temperature, has largest intensity in the
G-OO phase before being suppressed in the C-OO phase of YVO$_3$ due to the
specific low-temperature orbital ordering pattern. In their model of the
bi-orbiton, $d_{xz,1}$ and $d_{yz,2}$ orbitals on neighbouring sites 1 and 2
are excited to $d_{yz,1}$ and $d_{xz,2}$, conserving the spin order in the
process\cite{benckiser2008}. This is allowed in the G-OO/C-SO pattern, but
forbidden in the C-OO/G-SO pattern, explaining the temperature dependence
of the optical spectra. Our measurements reveal a similar behaviour with
temperature, in which feature `O' grows in intensity in the intermediate G-OO
phase, and persists into the G-OO/C-SO phase transition ($T_N = 138$~K). Below
$T_N$, however, recent high-resolution diffraction measurements have suggested
a coexistence of G-OO and C-OO for rare earth ions of intermediate size
(including NdVO$_3$)\cite{sage2006,sage2007}. The fraction of C-OO in this
phase-separated region is suggested to be as high 65\% for GdVO$_3$ and
TbVO$_3$\cite{sage2007}.
In La$_{1-x}$Lu$_x$VO$_3$, the suppression of G-OO in favour of C-OO (due to
cation disorder) has been found to result in large differences in the RIXS
intensities of the $^3T_2$ and $^1T_2$/$^1E$ excitations\cite{llvo}. In
Fig.~\ref{f:ndtemper}(a), the transition probabilities of these features are
observed to be notably different at 80~K compared with other temperatures,
consistent with a growing fraction of C-OO coexisting with G-OO below $T_N$ in
NdVO$_3$.
In the model of Ref.~\cite{benckiser2008}
this kind (C-type) of ordering would suppress the two-orbiton process, since it
consists of nearest neighbours with the same spin but also the same orbital
occupation. Even if the ratio of C-OO is low, the formation of small droplets
of C-OO within the G-OO matrix, as suggested by Ref.~\cite{sage2006},
would scatter the propagation of the delocalised orbital wave.

\begin{figure}[t!]
\begin{center}
\includegraphics[width=0.75\linewidth,clip]{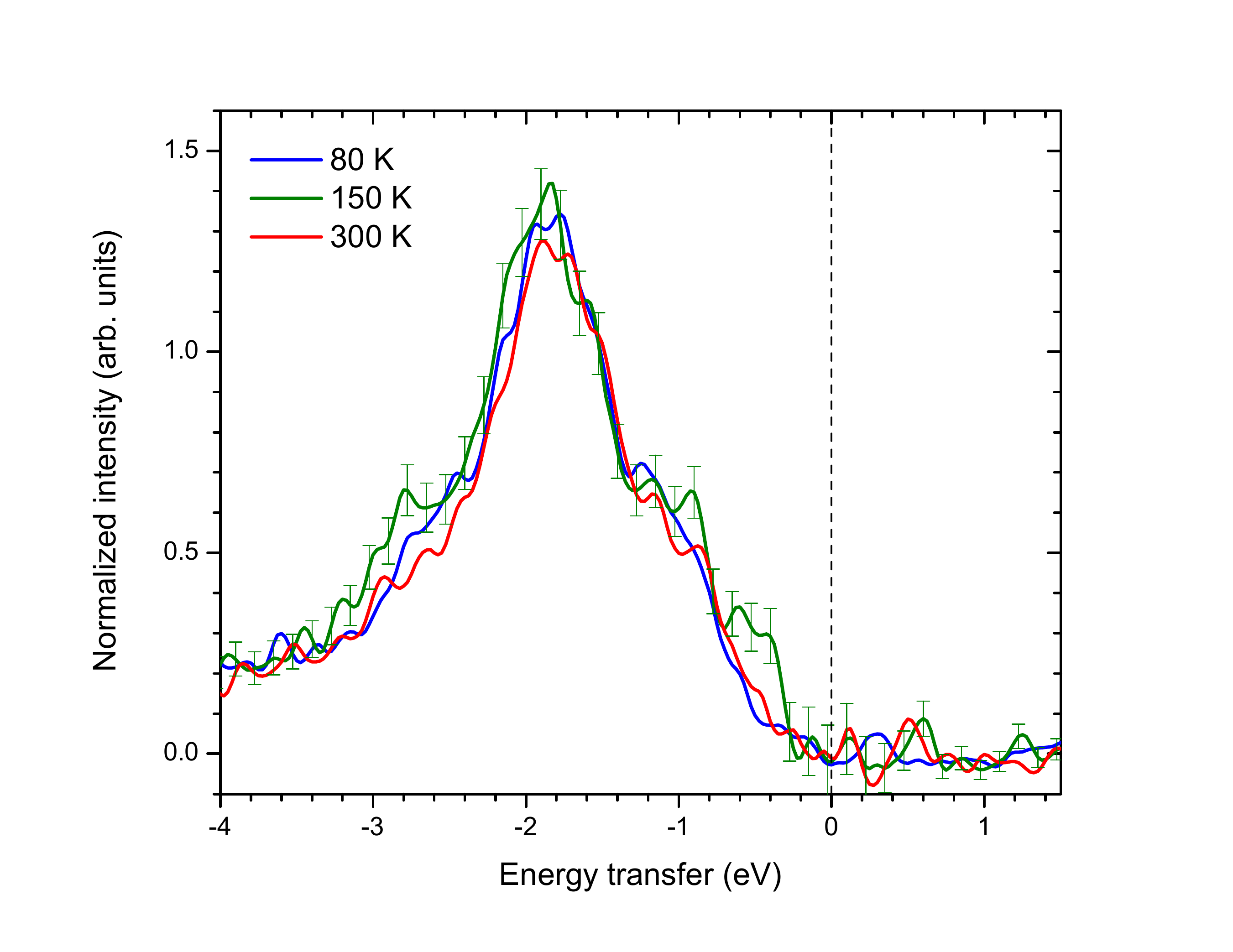}
\end{center}
\vspace*{-0.2in}
\caption{\label{f:ytemper} Temperature dependence of
RIXS excitations in YVO$_3$, recorded with an incident photon energy of
514.5~eV. The spectra are shown after removal of the elastic peak, in the same
way as Fig.~\ref{f:ndtemper}(a).}
\end{figure}

A very recent high-resolution
RIXS study of YVO$_3$ has also found a feature at 0.4~eV
in difference (between photon energies) spectra at the O $K$-edge, which
the authors link to the bi-orbiton\cite{benckiser2013}. However, although
their V $L_3$-edge RIXS spectra are in agreement with our CF energies, the
0.4~eV feature was not observed at the V $L$-edge in their data, eliminating
the possibility of directly associating this transition with V electrons. This
also raises the question of why the bi-orbiton feature is observed in our
NdVO$_3$ data at the V $L$-edge, but not by Ref.~\cite{benckiser2013} in
YVO$_3$ at
the same edge. In Fig.~\ref{f:ytemper}, we show similar temperature-dependent
V $L$-edge RIXS measurements at 514.5~eV on YVO$_3$, recorded with the same
instrument settings and treated in the same way as the NdVO$_3$ data
of Fig.~\ref{f:ndtemper}. Although some
intensity is observed near $-0.4$~eV, it is much weaker than in NdVO$_3$, and
its temperature dependence is only barely discernible. As pointed out by
Pavarini and co-workers in the rare-earth titanates, the precise details of the
electronic structure, and hence CF levels and orbital order patterns,
of perovskite-like materials is very sensitive to the {\em
A}-site hybridisation\cite{pavarini2005}. In particular, the hybridisation
between the unoccupied Y $4d$ and V $t_{2g}$ orbitals is likely quite different
to that between the broader and energetically higher lanthanide $5d$ bands and V
$t_{2g}$. Our measurements suggest that although YVO$_3$
is one of the most studied
{\em R}VO$_3$ compounds, it may not be representative of a typical
rare earth ion, and that future high-resolution studies of
orbital wave physics in these compounds may be more fruitful in other
compounds.

\section{Conclusions}
To summarise, we have presented a comprehensive RIXS study of the excitation
spectrum of NdVO$_3$ above and below the OO and SO transitions. Whereas most
of the experimental features show very weak temperature dependence and can
be explained very well from an atomic CFM perspective,
a low-energy feature is observed with an energy of $0.4$~eV. The
resonant behaviour of this peak confirms its origin as excitations of the V
$t_{2g}$ electrons. This feature cannot be explained as a local CF
excitation, and exhibits a strong temperature dependence consistent with
the orbital wave model of Ref.~\cite{benckiser2008}. In this model,
this peak is due to a collective bi-orbiton excitation, which is allowed in
the G-OO and G-OO/C-SO phases. Below $T^* \sim 125$~K, a finite fraction of
C-OO coexists with the G-OO, and suppresses the excitation and propagation
of the bi-orbiton.
Our measurements are consistent with this model, suggesting the $-0.4$~eV
feature observed in NdVO$_3$ may be associated with the bi-orbiton excitation.
This feature is found to be heavily suppressed in
similar RIXS measurements of YVO$_3$.
Future high-resolution and momentum-resolved RIXS experiments are encouraged to
unambiguously rule on its origin, and should be focussed near the $t_{2g}$
resonance of compounds other than YVO$_3$.

\section*{Acknowledgements}
The Boston University program is supported in part by the Department of Energy
under Grant No.\ DE-FG02-98ER45680. The ALS, Berkeley, is supported by the
U.S.\ Department of Energy under Contract No.\ DE-AC02-05CH11231. The NSLS,
Brookhaven, is supported by the U.S.\ Department of Energy under Contract
No.\ DE-AC02-98CH10886. G.B.\ gratefully acknowledges financial support from
EPSRC Grant EP/I007210/1.

\appendix
\section*{Appendix: Sub-Pixel Sampling}
The emission spectrometer used in these measurements is a Nordgren-type
Rowland circle
spectrometer \cite{nordgren1986}. The x-rays are collected from the sample
through an adjustable entrance slit assembly, after which they are diffracted by
a spherical grating.  The diffracted x-rays are subsequently detected by a 2D
multi-channel plate (MCP) detector. The image of the straight entrance slit
through the spherical grating is curved, necessitating the 2D detection system
(see, for example, Fig.~2 of Ref.~\cite{nordgren1986}). However, this curvature
also affords the possibility of sampling below the natural pixel width of the
MCP detector.

\begin{figure}[t!]
\begin{center}
\includegraphics[width=0.70\linewidth,clip]{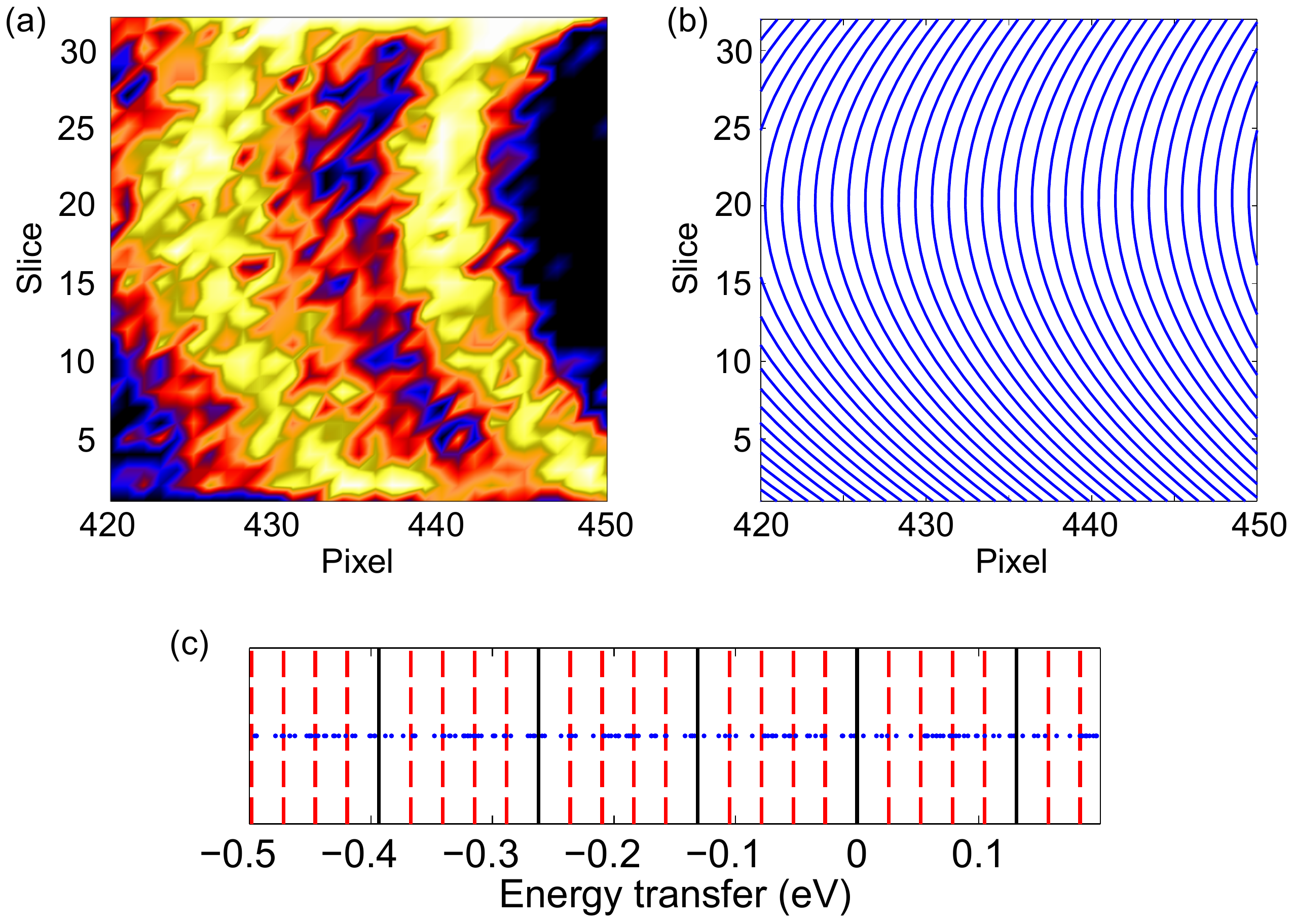}
\end{center}
\vspace*{-0.2in}
\caption{\label{f:mcp} (a)~Close-up of the image of the MCP of a
typical V $L_3$-edge RIXS measurement, illustrating the pronounced curvature in
the detected image. (b)~Iso-energy contours of the image shown in
(a), showing the curvature function.
(c)~Pixel coverage of the MCP shown in (a) and (b). Each dot represents the
centre in energy of one of the pixels in one of the slices in (a) and (b). The
solid vertical lines represent the natural pixel width of the MCP, and the
vertical dashed lines correspond to sub-pixel sampling at one-fifth of a pixel.}
\end{figure}

Shown in Fig.~\ref{f:mcp}(a) is the 2D image of the region near the elastic
peak of a representative RIXS measurement, displaying the curved
image of the elastic scattering, as well as a low-energy CF transition. The
2D image is binned into 1024 horizontal
``energy'' channels (actually evenly-spaced in wavelength), and 32 vertical
``slices''. Next to this image, in Fig.~\ref{f:mcp}(b),
iso-energy curves are shown for the same measurement at the same part of the
detector. In
addition to the obvious offset in energy between each slice, the
individual slices each have their own (different) calibration (pixel width). It
is clear that, in general,
there is a non-integer offset (in pixels) between each slice, which means
that the centres of each pixel do no align. When correcting for the curvature,
one is therefore faced with two possibilities: 1) to share the counts of a pixel
that is offset by a fractional amount with the bins either side (i.e.~bilinear
interpolation), which is the conventional approach, or 2) to sample the data
below the natural pixel width, and take advantage of this extra fidelity.

Sub-pixel sampling does not improve the instrument resolution, of
course, since this is limited by the optics of the detector and the width of the
entrance slit. However, it does allow for higher data fidelity, which is
useful in assessing whether low-energy features of the data are due to random
noise, or intrinsic to the system under investigation. At broad resolution
functions, there is little to be gained by sub-pixel sampling, since one is
already sampling at a small fraction of the resolution function using the
natural pixel width. However, in our measurements, we set the instrument
resolution function to 0.26~eV FWHM (the smallest achievable slit width with
manageable count rates), and the natural pixel width of 0.13~eV only samples the
data at half of the resolution FWHM (or more than 1 sigma). By using sub-pixel
sampling, we are able to increase the fidelity to one tenth of the FWHM (or one
quarter of sigma). In Fig.~\ref{f:mcp}(c), the centre in relative energy of
each pixel is shown by the points, with the conventional natural pixel width
indicated by the vertical solid lines, indicating the energy coverage of the MCP
detector. The vertical dashed lines correspond to sub-pixel sampling every fifth
of a pixel, with many
pixels contributing to each sampling point.

Owing to the varying pixel coverage of sub-pixel sampling, and the unknown
errors introduced by the (bilinear) interpolation, a reliable error map is
difficult to quantify at the sub-pixel level. Instead, in Figs.~4 and 5 of the
manuscript, we have chosen to present the data using statistical error bars
for the data points that are sampled at the natural pixel width, with the
sub-pixel sampled data points shown as lines connecting these points (in effect,
`guides for the eye'). We believe
this is the fairest way of the presenting this kind of analysis, where we make
no direct claim on the error distribution of the sub-pixel sampled data points,
but this information is provided as a qualitative impression of the higher
fidelity that is available.
Finally, the analysis of simulated data indicates that this approach does not
introduce artefacts in the data.
We emphasise that no cubic interpolants, or other
high-order interpolation methods are used -- only bilinear interpolation
is used to distribute the counts (which
is incapable, by itself, of developing peaks between data points).

\end{document}